# A Low-noise Germanium Ionization Spectrometer for Low-background Science

Craig E. Aalseth, Juan I. Collar, Jim Colaresi, James E. Fast, Todd W. Hossbach, John L. Orrell,
Cory T. Overman, Bjorn Scholz, Brent A. VanDevender, K. Michael Yocum

*Abstract*—Recent progress on the development of very low noise high purity germanium ionization spectrometers has produced an instrument of 1.2 kg mass and excellent noise performance. The detector was installed in a low-background cryostat intended for use in a direct detection search for low mass, WIMP dark matter. This Transaction reports the thermal characterization of the cryostat, specifications of the newly prepared 1.2 kg p-type point contact germanium detector, and the spectroscopic performance of the integrated system. The integrated detector and low background cryostat achieved full-width-at-half-maximum noise performance of 98 eV for an electronic pulse generator peak and 1.9 keV for the 1332 keV $^{60}$Co gamma ray.

## I. INTRODUCTION

DEVELOPMENT of p-type point contact (PPC) high purity germanium ionization spectrometers for application to low-energy threshold measurement of ultra-rare nuclear and particle physics processes was accelerated in the mid-2000s [1]. Initial detectors were employed to pursue measurement of coherent scattering interactions of reactor-produced antineutrinos with germanium nuclei [2] and later to search for direct detection of dark matter particles [3]-[5]. In both cases one seeks to measure the ionization generated by the recoil of a germanium nucleus after being struck by a low energy antineutrino or dark matter particle. As of the preparation of this Transaction, no confirmed detection of either neutrino-nucleus coherent scattering or direct detection of dark matter has been produced, either using PPC germanium detectors or any other detection method under development. The experimental challenge in all such detection systems is the two-fold problem of seeking a signal that is just below both the energy threshold of the detector and the intrinsic background of the instrument.

This Transaction reports further development of the PPC germanium detector technology. Specifically, initial generations of this class of high purity germanium detector were limited to less than 0.5 kg total mass and an analysis energy threshold of ~0.5 keV. As reported previously, a low-background cryostat was developed to accept a large (~1 kg) germanium crystal, specifically targeting a reduction in the achievable analysis energy threshold through improved electronic noise performance [6]. In parallel, CANBERRA Industries undertook a sustained program to produce (and reproduce) a PPC germanium detector design able to achieve noise performance of less than 100 eV full-width-at-half-maximum (FWHM) from an input electronic pulse generator, down from the ~160 eV FWHM achieved by the previous generation of PPCs. This Transaction reports the final preparation of the cryostat, initial laboratory performance of a 1.2 kg p-type contact germanium detector, and preliminary results after the integration of these two pieces of equipment.

## II. CRYOSTAT CONSIDERATIONS

The design and aims of the low-background cryostat under consideration were presented in a prior Transaction [6]. A photograph of the finished cryostat is shown in Fig. 1. To summarize those considerations, the low-background cryostat was designed to support large diameter (up to 90 mm) germanium crystals in an open, crystal-mount design intended to reduce parallel capacitance that can contribute to the electronic noise of the detector system. The cryostat followed low-background design principles using a convenient combination of ultra-pure electroformed copper [7] and commercial stock OFHC copper. Attention to removing all outer layers of stock OFHC copper material, followed by etching and passivating of all surfaces, was employed in the background reduction process [8]. The amount of plastic (PTFE) was minimized both to reduce possible contributions to the detector system's electronic noise via dielectric loss as well as to reduce the background produced by any naturally occurring radioactive isotopes entrained as minor impurities in the plastic material. The vast majority of handling and assembly of the cryostat and crystal mount components took place in a clean room in Pacific Northwest National Laboratory's shallow underground laboratory [9].

Although these low-background design and fabrication principles were followed, the cryostat does not represent the absolute lowest background assembly achievable. For example, the instrument is not composed entirely of ultra-pure electroformed copper, nor has deliberate attention been given to minimizing the cosmic ray exposure of the cryostat or crystal mount components.

Manuscript received March 3, 2016. Please see the Acknowledgments for author affiliations and agencies that provided support for this work. The corresponding author is J. L. Orrell (phone: 1-509-375-1899; fax: 1-509-372-4591; e-mail: john.orrell@pnnl.gov).

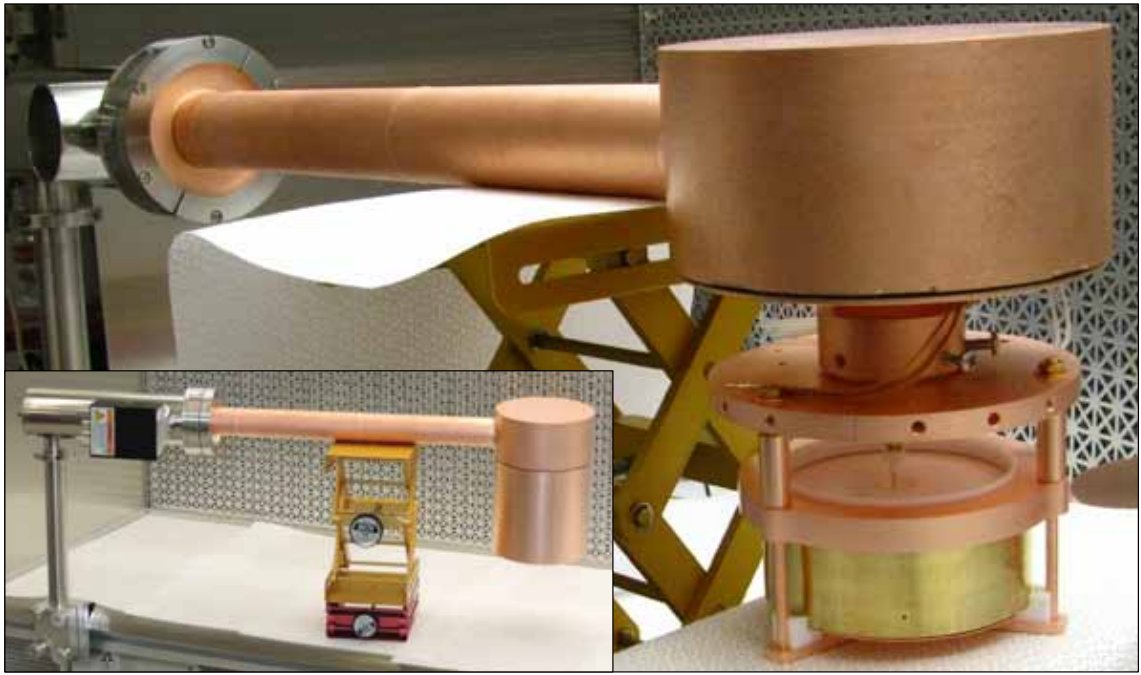

Fig. 1. Photograph of the finished cryostat described in [6] and reported on in this Transaction. In the larger picture, the outer vacuum jacket and infrared shield are removed to display the open crystal-mount design and a brass "crystal blank" put in place of the germanium crystal for assembly and testing of the cryostat. The crystal blank is the correct mass (*i.e.* 1210 grams) to simulate the anticipated mass of germanium in the final configuration. The approximately 20% higher heat capacity of brass compared to germanium (at room temperature) means the expected cool-down time will be longer. The inset photograph shows the fully assembled Al dipstick, black preamplifier box, copper cross-arm, and copper end-cap, supported by a pair of lab-jacks sitting in a Laminar flow hood.

Separate work [10] evaluated the anticipated backgrounds of the cryostat discussed in this Transaction in the case of an underground deployment to search for low-mass dark matter and concluded the principal background contributions would not be due to the cryostat fabrication materials. However, if it is later found the cryostat materials contribute significantly to the residual background of the instrument, future generations of the cryostat design presented in this Transaction could reach ultra-low-background levels simply through more attention to selection and fabrication of materials (*i.e.*, no redesign of the cryostat is required).

*A. Cryostat Performance*

Between periods when the cryostat was opened for work, the pressure was measured to be $2 \times 10^{-7}$ Torr during long periods of vacuum pumping. This value is typical of "good vacuum" in commercial germanium detector cryostats. This vacuum pressure was measured with the following seals: a commercial double Viton o-ring seal for the vacuum tube connection, one single indium o-ring seal and one single Viton o-ring seal at the connection between the commercial liquid nitrogen "dipstick" and the low-background cryostat, and one single Viton o-ring seal at the copper end-cap. All other vacuum connections were made using standard ultra-high-vacuum commercial copper gaskets set in knife-edge, bolted vacuum flanges. During the final assembly an indium o-ring seal was used at the copper end-cap rather than a Viton o-ring. The dipstick and cryostat were continuously pumped-on or kept under vacuum in an attempt to maintain the maximum absorbency of the molecular sieve employed as a cryogenic getter during cold operation.

In preparation for the cool-down test of the cryostat and brass "crystal blank", a cart was assembled to hold a standard open-neck 30 L liquid nitrogen (LN) dewar (*e.g.*, CANBERRA Model D-30/OS offset neck dewar) and the full cryostat assembly. To facilitate temperature measurements, the electroformed copper end-cap was replaced with an Al end-cap that included an electrical feed-through. This electrical feed-through was used to instrument and read-out the interior of the cryostat with four LakeShore DT-670 Si diode temperature sensors (CU package). The sensors were connected to a Lakeshore 218 Temperature Monitor unit to provide the temperature read-out. Three of the DT-670 sensors were mounted at the cold-end of the Horizontal Cold Finger (HCF), the middle of the Vertical Cold Finger (VCF), and the top of the Cold Plate (CP), as shown in Fig. 4. The fourth DT-670 sensor was mounted inside the brass crystal blank, as shown in Fig. 2, with a copper foil used to obscure the line of sight from the IR shield copper shield to the diode. The goal for the fourth DT-670 is to obtain a representative measurement of the bulk temperature of the crystal blank.

The cryostat cool-down test was a simple affair of filling the LN dewar with liquid nitrogen (LN) and recording the cool down of the system via the four LakeShore DT-670 temperature sensors. A Mettler scale, upon which the entire cart, cryostat, and dewar were placed, was used to measure the rate of LN boil-off, giving a measure of the total heat load of the dewar and cryostat system as well as tracking when LN refills were required. The cool-down test data are presented in Fig. 3.

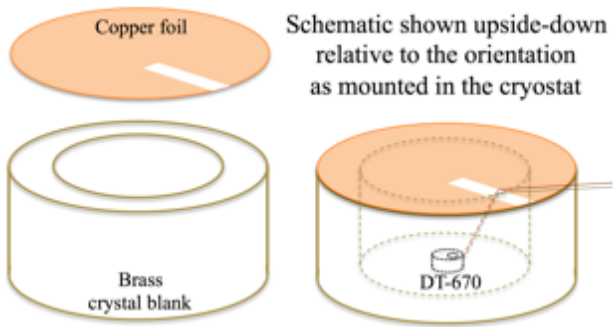

Fig. 2. A schematic representation of the brass "crystal blank", LakeShore DT-670 temperature sensor mounting location, and IR-blocking copper foil.

After the cryostat cool-down data was collected, it was recognized there is sufficient precision in the temperature sensor data to determine not only the copper-to-copper joint resistances but also transient thermal loading via IR due to changes in the room temperature of the lab where the cryostat was operated. A second data set was collected for all four temperature sensors, each independently, after they had been removed from the cryostat. They were placed on a copper rod and immersed in LN. This second data set showed the sensors varied in measured absolute temperature for the LN bath, reporting temperatures ranging from 77.33 K to 77.44 K. The specification sheets for the sensors imply an absolute temperature accuracy of only approximately 1 K. One of the sensors (sensor #2) was used to calculate an offset temperature ($\Delta T_{\#i,\text{offset}} = \Delta T_{\#i} - \Delta T_{\#2}$) for each of the other sensors, #$i$, from the second data set. Subtracting sensor #2's temperature value from the other three sensors' temperature values also removed correlated variations over time in the reported temperature of the LN bath from the three sensors having temperature offsets. This collection of temperature-offset data showed at ~77 K the remaining point-to-point variation of the three sensors ranged from 0.0017 K < $\sigma_T$ < 0.0019 K (or 4-8% calculated as $\sigma_T/\Delta T_{\#i,\text{offset}}$) where $\sigma_T$ is the square root of the sample variance. This is a surprisingly precise point-to-point temperature recording. Thus the sensors appear to have significantly finer precision than absolute temperature accuracy and this precision is used in studying the *relative variations* observed in the collected cryostat cool-down data, keeping in mind the manufacturer's specified limit of accuracy of ~1 K.

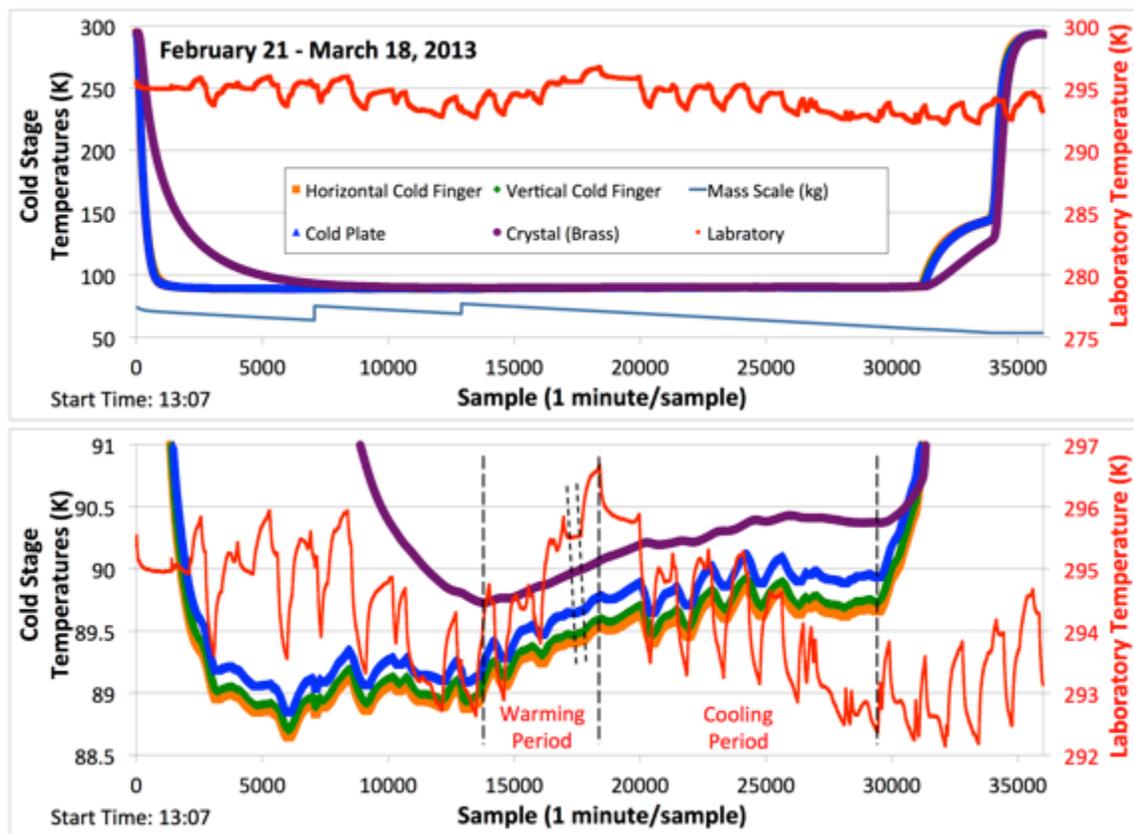

Fig. 3. Plots of temperature vs. time for a cool-down and warm-up cycle of the cryostat when the brass "crystal blank" was installed and the cryostat was instrumented with temperature sensors. The upper panel shows the full temperature range from room temperature to near liquid nitrogen temperature. The cold-stage temperatures (left vertical axis) over-lay each other in this figure except for the brass crystal blank trace (purple), which does not reach below 100 K until after the 5000[th] sample point. The room temperature sensor (right vertical axis) is shown over a smaller (1/10[th]) temperature range and appears as the jagged undulating trace (red) along the top of the upper panel. Also shown in the upper panel – the bottom-most saw-tooth trace (blue-grey) – is the mass reported by a scale, essentially tracking the boil-off and refilling of the liquid nitrogen dewar used to cool the cryostat (use the left vertical scale, but read the numerical values as kilograms rather than Kelvin). The lower panel shows a finer scale for both the cold stage temperatures near liquid nitrogen temperature (left vertical axis) as well as the room temperature sensor (right vertical axis). The cold stage temperature traces listed from top to bottom (warmest to coldest) are the brass crystal blank (purple), cold plate (blue), vertical cold finger (green), and horizontal cold finger (orange). The variations in the room temperature trace (red) clearly impact temperature reported by the temperature sensors. See the text for a discussion of the extent to which this is an artifact of the sensor placement and/or representative of the actual forcing of the cold stages by the room temperature surroundings.

## B. Thermal Model

In a prior publication [6] on the development of the cryostat reported in this Transaction, a simple one-dimensional, electrical-resistance-analog model calculation was used to estimate the anticipated thermal performance of the cryostat. In the prior publication [6], the germanium crystal operating temperature was estimated to be 103 K based on input parameters to the (simple) thermal model including estimates for the electrical (0.07 W), conductive (0.5 W), and radiative (2.2 W) heat loads, which included joule heating from the front-end electronics, conduction through the thermal stand-offs, high-voltage wiring and signal wiring, and thermal radiation from the vacuum jacket. The radiative-load calculation used $\varepsilon = 0.03$ for the room-temperature (300 K) copper emissivity [11], [12], and due to the freezing-out of water vapor on the cold surfaces, assumes a value of $\varepsilon = 1$ for the cold-side emissivity.

To improve upon the simple thermal model that was based on simplified geometries for the radiative loading [6], a steady state COMSOL finite element model was developed. The COMSOL model is composed of the as-built cryostat geometry from the dipstick connection to the crystal infrared shield and end can. To reduce the complexity of the FEM mesh, the crystal, crystal support structure, FET assembly, threaded holes, and thermal stand-offs were removed from the model. The experimental temperature sensor footprints were included in the geometry to compare model temperatures to the experimental temperatures, shown in Fig. 4.

Estimates of the copper thermal conductance curve [13] and of the residual heat loads from wiring and thermal standoffs are used as inputs to the thermal model. The dipstick connection surface is assigned a temperature of 88.63 K using the following relation:

$$T_{\text{dipstick}} = 82.42 \text{ K} + 5.74 \frac{\text{K}}{\text{W}} \left( P_{\text{cryostat power}} \right)$$

This relation was determined from laboratory testing, where the cryostat power consumption was calculated to be 1.08 W. The emissivity of copper has low temperature dependence between 50 K and 300 K [14], therefore the emissivities of the cold and warm surfaces were set equal to each other, but allowed to float as a single free parameter for model tuning. The joint resistances were initially set to zero to model temperature sensor differentials due to the bulk resistivities of the cryostat. Along with temperature differentials, the average heat flow through each joint can be calculated to determine the bulk resistivity due to conductive losses.

The joint resistances, $R$, were calculated to be 0.01 K/W (HCF/VCF) and 0.21 K/W (VCF/CP), based upon the temperature difference across the joint divided by power, $P$, using equation

$$R_{\text{joint}} = \frac{\Delta T_{\text{experiment}} - \Delta T_{\text{joint model}}}{P_{\text{joint model}}}$$

taking into account the temperature sensor offsets during a relatively stable period of the room temperature. The temperature offsets are included by having $\Delta T_{\text{experiment}} = (T_{\#j} - \Delta T_{\#j,\text{offset}}) - (T_{\#i} - \Delta T_{\#i,\text{offset}})$, for two different sensors, #i and #j, on either side of a joint. The stable period of the room temperature (red line) is identified in Fig. 3 as the period between the two diagonal dashed lines during the warming period.

Several iterations of the COMSOL model were performed to select a best match between the model temperatures and the measured (offset corrected) temperature data. In particular the COMSOL model gave the thermal joint resistance estimates based on matching the *temperature differences* between the model components (*i.e.*, HCF, VCF, and CP) while the copper emissivity estimates were based on the *absolute temperatures* measured. For the latter, an emissivity of 3.2% was determined as a best match to the temperature data, which is close to the estimated value in the simple thermal model of [6] and matches previous reports [11], [12].

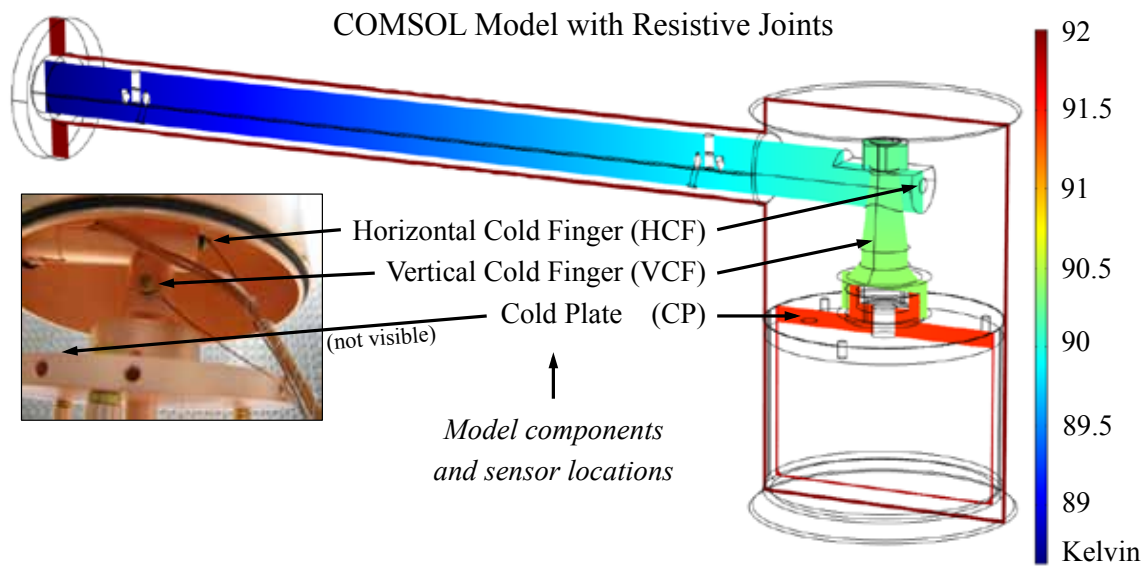

Fig. 4. Visualization of the COMSOL model used to estimate the copper-to-copper thermal joint resistances and emissivity of the copper surfaces within the cryostat. The inset photograph shows via arrows the locations of three of the mounted LakeShore DT-670 temperature sensors. See text for further description.

The results for the joint resistance estimates between the simple thermal model and the COMSOL thermal model are presented in Table I. Overall the final COMSOL thermal model, shown in Fig. 4, comes to within 1 K of the measured absolute temperatures in the cryostat system when held near LN temperature. The accuracy level of the sensors (±1 K) must be considered. With an agreement of ±1 K between the model and the collected data, the thermal modeling effort was halted as the level of uncertainty is comparable to the uncertainty associated with the temperature sensors. Furthermore, during the COMSOL model iterations it became apparent a more detailed model of the dipstick portion of the cryostat was needed to account for the dewar heat load, rather than using a fixed, infinite thermal heat bath representation. In summary, the conclusion is the COMSOL thermal model matches the data to within the sensor accuracy.

TABLE I
COMPARISON OF THERMAL MODEL JOINT RESISTANCE ESTIMATES

| Joint Resistance | Simple Model [6] | COMSOL Model |
|---|---|---|
| $R_{HCF/VCF}$ | 0.96 K/W | 0.01 K/W |
| $R_{VCF/CP}$ | 1.82 K/W | 0.21 K/W |

It is natural to question if the *time-dependent* variations in the room temperature are truly changing the temperature of the cryogenic system, as seems apparent by the reported temperature sensor data shown in Fig. 3. Using the COMSOL thermal model, two static calculations with the room temperature offset by 1 K were performed. The results from the COMSOL model imply the temperature of the cryogenic system should change by only 0.02 K on the HCF and 0.04 K on the cold plate. These numbers are in stark disagreement with the ~0.5 K temperature swings shown by the HCF, VCF, and CP sensors in the cool-down data of Fig. 3. Referring to the manual for the sensors, one is cautioned that the sensors must be fully enclosed in a constant temperature environment to give an accurate temperature measurement. Three sensors have line of sight to the warm outer vacuum jacket, a situation specifically noted as problematic in the sensor manual. However, the crystal sensor is correctly fully enclosed. Considering the span of time between the vertical dashed lines labeled "warming period" in Fig. 3, it is possible to infer that the crystal temperature rises 0.22 K per 1 K change in room temperature. However, this is still inconsistent with the COMSOL model. Finally, the last qualitative evaluation of the temperature data is an observation of the lag-time for the room temperature change to affect the temperature reported by the temperature sensors, of about 5 hours, as shown by the pair of diagonal dashed lines shown in Fig. 3.

Although the time variation of the collected temperature data is very interesting from an academic stand point, it is noted that in the detector's planned operating environment – buried deep within a lead shield – should remain steady and variations such as seen in Fig. 3 will not be induced, thereby ensuring long-term stability of the germanium detector temperature. Nevertheless, these observations suggest future improvements should include use of sufficiently low radioactive content super-insulation or floating IR shields to provide better cold-stage stability. Maintaining long-term stable operation is particularly relevant when searching for an annual modulation of WIMP dark matter interactions [15].

### C. Plastic Dielectric-loss Noise Study

As part of the development of the cryostat, an investigation was performed in an attempt to determine the extent to which dielectric materials play a role in the increase in electronic noise of a detector system due to dielectric loss. A special Low-Energy Germanium (LEGe) detector system was prepared by CANBERRA based on specifications to include a volume of dielectric material many times larger than normally required to insulate the detector high voltage surfaces from the cryostat mounting. LEGe-style detectors have characteristics similar to those of PPC detectors, namely low-energy performance in the several keV x-ray energy range, low-noise, low-capacitance, and similar electronic (high voltage and read-out) configuration. Energy resolution was measured as 153 eV at 5.9 keV from an $^{55}$Fe radioactive source. Noise performance of the detector and AmpTek PX5 HPGe system would permit ~0.3 keV energy thresholds. A Maxwell 3-D [16] model showed the capacitance to the read-out pin was ~2 pF in comparison to the typical 2-3 pF capacitance of large volume PPC detectors. Thus the LEGe detector provides a suitable test configuration for investigating the potential impact of dielectric materials used within the cryostat housing to mount and electrically insulate germanium ionization spectrometers.

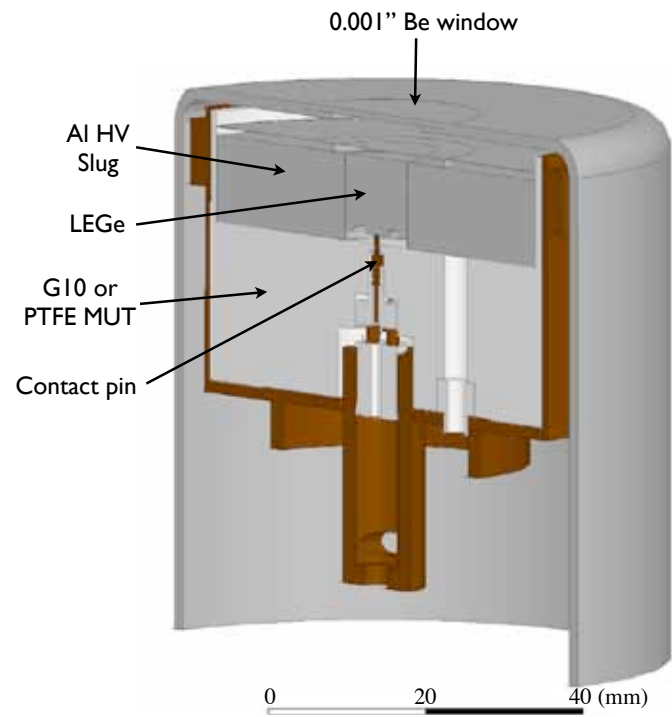

Fig. 5. Diagram of the special LEGe detector design intended to test the impact of a large amount of dielectric material on noise performance of low-noise germanium detectors. The entire aluminum slug is raised to the high voltage required to operate the LEGe detector. The G10 or PTFE material under test (MUT) is shown surrounding the read-out contact pin and electronics front-end package.

Measurements consisted of using a digital multi-channel analyzer (AmpTek PX5-HPGe) to collect energy spectra from the LEGe detector using an $^{55}$Fe radiation source, an AmpTek Cool-X x-ray generator that emits Cu x-rays, and an digital pulse generator to place six peaks in the 2-10 keV range (calibrated via the above noted x-ray sources). The average energy resolution ($\langle\sigma\rangle$) of the six electronic pulser peaks gives the electronic noise contribution to the detection system. The width of the electronic pulser peak will vary as the shaping time of the energy-calculating filter in the PX5-HPGe changes. Plotting the average electronic pulser energy resolution widths squared ($\langle\sigma\rangle^2$) as a function shaping time on a log-log figure shows a roughly parabolic form, from which the parallel, $1/f$, and $f$ noise component contributions can be compared. These tests were done for G10 and PTFE test samples. While the relative decomposition into the three noise components did differ for the PTFE and G10 test samples, the actual measured best electronic pulser peak resolutions (minimum $\sigma = \sim45$ eV) were indistinguishable to within the ~10% error of the measurement results.

In summary, the simplistic test described in this sub-section was intended to pose a worst case scenario of a large volume of "poor" dielectric material (G10) placed in experimental comparison to a "good" dielectric material (PTFE), showed no appreciable difference in ultimate detector performance. Although this test does not address the issue of the dielectric material used in the mounting substrate for any discrete front-end electronic components, it is believed this test does provide circumstantial evidence that noise will not be introduced into the detector system due to the amount or choice of dielectric material used to hold and mount germanium detectors.

## III. Detector Characteristics

Table II reports the physical characteristics of the high purity germanium crystal selected for the low-noise detector development, photographed in Fig. 6. This p-type point contact (PPC) germanium detector was selected, prepared, and tested at CANBERRA in Meriden, CT. The results from those efforts are reported as the expected baseline performance in a standard CANBERRA test cryostat. The detector was then mounted in the low-background cryostat described in the previous sections of this Transaction and a new set of performance metrics were collected for comparison. Finally, the detector and cryostat were transported to the University of Chicago for further evaluation.

TABLE II
DETECTOR PHYSICAL CHARACTERISTICS

| Parameter | Value | Unit |
|---|---|---|
| Diameter | 92.1 | mm |
| Height | 36.6 | mm |
| Weight | 1268 | g |
| Active Volume | 223.7 | cc |
| Li Dead Layer | 0.7 | mm |
| Point Contact Diameter | 3.5 | mm |

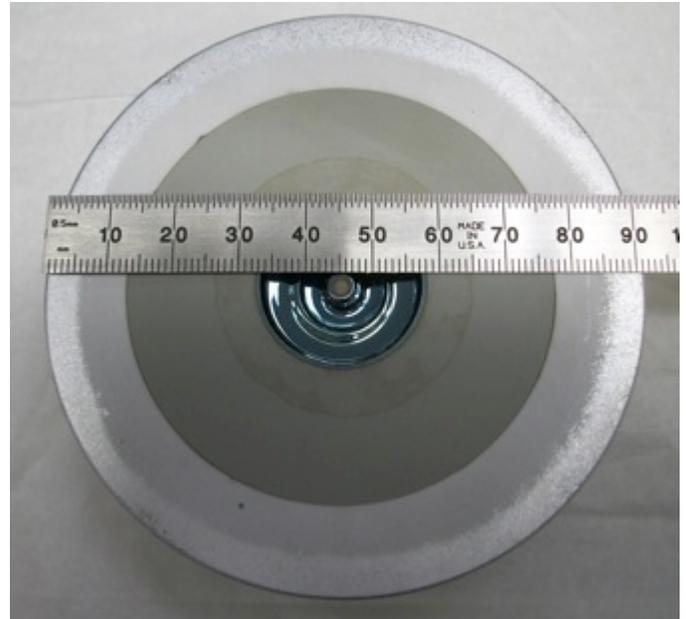

Fig. 6. Photograph of the p-type point contact detector (P3961AA) described in Table II showing the boron implanted P+ contact in the center with a gold ohmic metallization, the annular passivated junction surface, and the N+ lithium contact that encompasses the remainder of the surface outside the junction region. The lighter ring at the outer periphery is the sputtered aluminum ohmic contact that facilitates mechanical contact to the lithium.

### A. Initial Detector Selection & Performance

The initial detector performance reported in this sub-section was conducted at CANBERRA in Meriden, CT. Extensive investigation of the various components related to achieving the lowest noise in a large volume detector were carried out prior to fabricating the final 90 mm PPC detector. First, field modeling studies were performed using QuickField$^{TM}$ finite element analysis (FEA) software to determine the best device geometry and impurity distribution for the 90 mm geometry, as seen in Fig. 7. Through the FEA modeling, a cylindrical geometry with a wraparound lithium contact was determined to be preferable to bulletized or open-face geometries. Furthermore, an impurity concentration range from $7.0\times10^9$ cm$^{-3}$ to $9.3\times10^9$ cm$^{-3}$ was chosen to give the required depletion and charge collection performance.

Junction field effect transistors (JFETs) from several sources were tested and the Moxtek MX-20 was chosen as best for the selected detector geometry. Investigations were carried out to determine the best junction surface treatment (*i.e.*, contact surface passivation) to use on the point contact. The study compared amorphous hydrogenated silicon (aHSi), amorphous hydrogenated germanium (aHGe) and silicon oxide (SiOx). Two different ~570 gram detector crystals were used to investigate the performance as a function of the junction surface treatment. The aHSi proved to be the least suitable treatment and the aHGe and SiOx gave essentially equal performance, as determined by the measured full-width-at-half-maximum resolution of an electronic pulser induced signal. SiOx was chosen as the treatment to use on the final 90 mm PPC detector crystal.

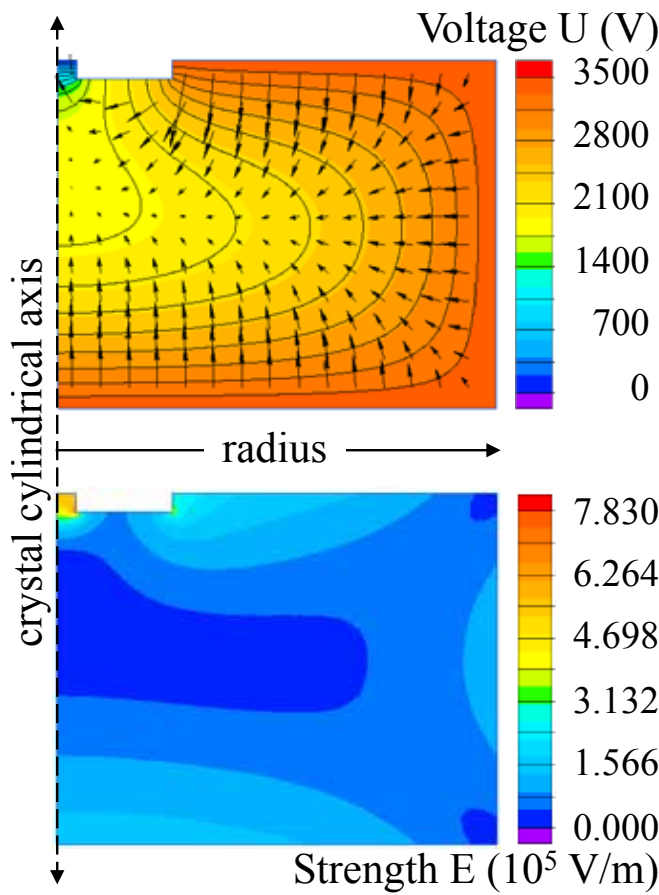

Fig. 7. Top: Simulation of equipotential distribution (200 V spacing) for the 90 mm PPC detector at bias voltage of 3500 V. Bottom: Contour plot of field strength distributions calculated for the same detector. The groove dimension is 4 mm inner diameter × 23 mm outer diameter × 2 mm depth. Only one-half of the detector cross-section is displayed.

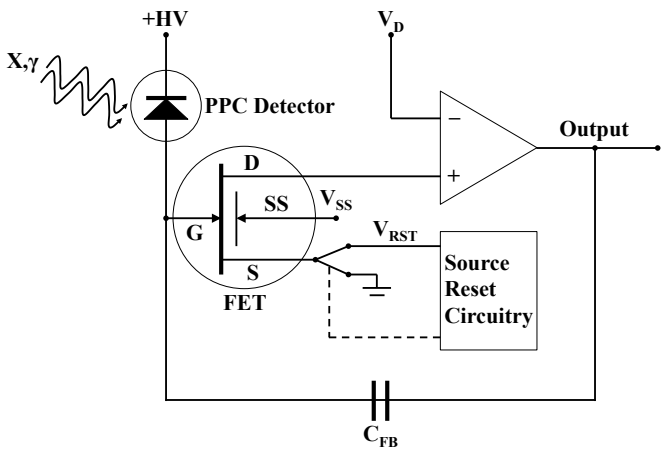

Fig. 8. Model DPRP with patented (Patent # US 6,587,003 B2) source reset mechanism. The current through the p-type point contact detector is integrated over the capacitor $C_{FB}$ until the output reaches a threshold and a reset through the source of the JFET is triggered.

The germanium detector crystal shown in Fig. 6 was initially housed in a commercial dipstick cryostat using a standard CANBERRA Dual Polarity Reset Preamplifier (DPRP) and a Moxtek MX-20 cooled JFET as shown in the block diagram in Fig. 8. The JFET receptacle inside the test cryostat housed a 1-kΩ heater resistor that could be used to adjust the operating temperature of the JFET, which typically gives best noise performance in the range of 140-160 K. The same JFET used in the test cryostat was ultimately used in the low background cryostat described earlier in this report. A heater resistor was *not* used in the low background cryostat due to the proximity to the detector and the undesirable amount of U & Th introduced with the addition of the heater resistor [3].

The detector operating characteristics, when housed in the CANBERRA test cryostat, are listed in Table III. The temperature of the crystal holder was measured in the test cryostat as 91 K.

TABLE III
DETECTOR OPERATING CHARACTERISTICS
CANBERRA TEST CRYOSTAT

| | |
|---|---|
| Full depletion voltage | 2700 V |
| Operating voltage ($V_{op}$) | 3500 V |
| Maximum operating voltage | 4000 V |
| Leakage current at $V_{op}$ | < 4.5 pA |

At the operating voltage ($V_{op}$), the detector was tested with a number of gamma-ray sources and evaluated for a standard set of performance metrics presented in Table IV (all energy resolution measurements reported were made with an 8 μs analog shaping time). The electronics chain comprised the following Canberra instrumentation: Preamplifier Model DPRP, Spectroscopy Amplifier Model 2026X-2, HV Power Supply Model 3106D, Electronic Reference Pulser Model 1407M, MCA Model Multiport II. For each isotope the full-width-at-half-maximum (FWHM) and full-width-at-tenth-maximum (FWTM) is reported for the specified gamma-ray energy peak.

TABLE IV
Detector Performance
CANBERRA Test Cryostat

| Isotope | $E_\gamma$ (keV) | FWHM (eV) | FWTM (eV) |
|---|---|---|---|
| $^{241}$Am | 59.4 | 331 | 607 |
| $^{57}$Co | 122 | 472 | 865 |
| $^{60}$Co | 1332 | 1823 | 3406 |

An electronic reference pulser was used to evaluate the noise performance and optimal shaping time for the detector system. Data was collected to determine the FWHM of the pulser peak for analog shaping times of 1, 2, 4, 8, 12, and 24 μs. In the CANBERRA dipstick test cryostat, with no heat applied to the JFET, the 12 μs shaping time gave 82 eV FWHM noise pulser peak width. With the heater on and the JFET temperature optimized, the lowest noise achieved with this detector in the test cryostat gave a pulser resolution of 60 eV FWHM at 12 μs shaping time as shown in the noise plot in Fig. 9. This is a dramatic improvement with respect to the 160 eV FWHM of the previous generation of PPCs [3], especially when taking into consideration the increase in active mass of the crystal by almost a factor of four.

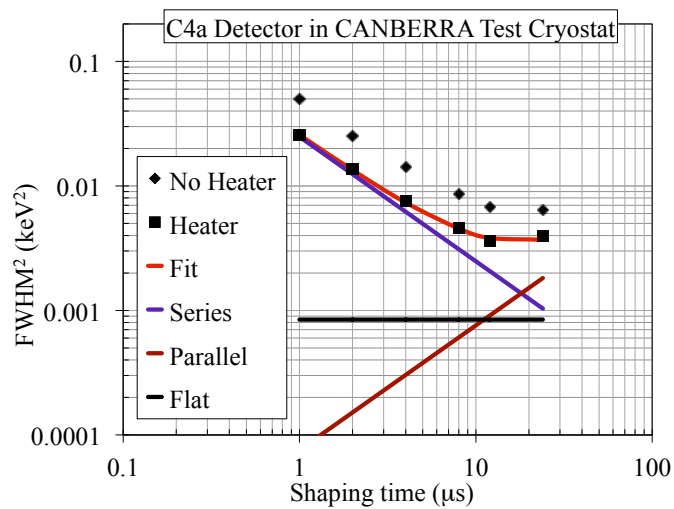

Fig. 9. Noise performance of the 90 mm p-type point contact detector ("C4a") and MX-20 JFET when tested in the dipstick test cryostat with optimized JFET parameters, external lead shielding, and controlled ambient noise. Measurements are made with and without a heater resistor to warm the JFET.

### B. Cryostat-Detector Integration & Performance

The detector performance reported in this sub-section was determined after the crystal was integrated into the low-background cryostat described earlier in this Transaction. The integration and subsequent performance characterization were performed at CANBERRA in Meriden, CT. A DPRP pulse reset preamp was used in the collection of the results presented in this sub-section.

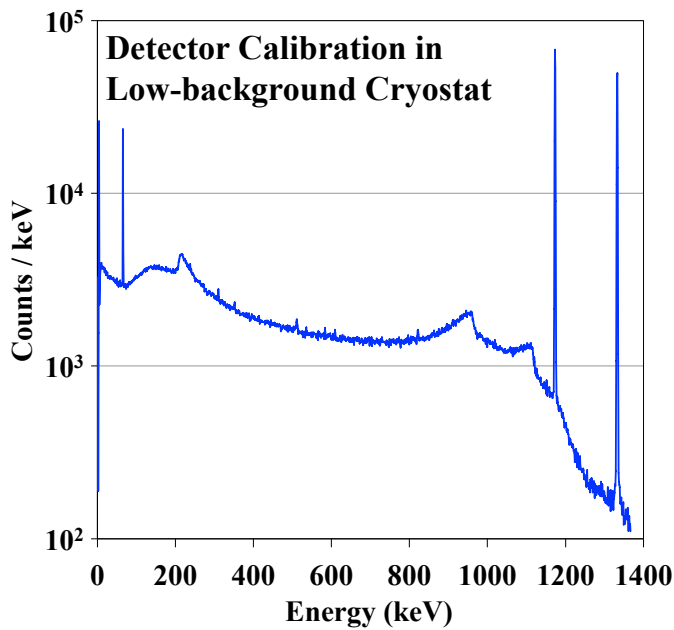

Fig. 10. Calibration of the point contact detector in the low-background cryostat using a $^{60}$Co source (gamma-ray peaks at 1173 keV and 1332 keV) and an electronic pulser (at approximately 65 keV).

Fig. 10 shows the energy spectrum collected using a $^{60}$Co gamma-ray source; an electronic pulse generator peak is also present in the spectrum at approximately 65 keV. The energy threshold was not determined for this system owing to the triggering challenge associated with any near threshold measurement. The energy threshold will be determined with a tuned data acquisition system at the underground location of the intended dark matter search.

The characterization performed in the low background cryostat has the same operating parameters previously presented in Table III. The energy resolution performance in the low-background cryostat (using 8 μs analog shaping time) is presented in Table V, for comparison to the performance obtained in the CANBERRA test cryostat (see Tab. IV). The noise performance in the low-background cryostat is measured in the same manner as described in the previous sub-section with results presented in Fig. 11, the best performance being at 8 μs shaping time with a FWHM of 98 eV *without* the use of a heater resistor to warm the MX-20 JFET.

TABLE V
Detector Performance
Low-background Cryostat

| Isotope | $E_\gamma$ (keV) | FWHM (eV) | FWTM (eV) |
|---|---|---|---|
| $^{241}$Am | 59.4 | 342 | 628 |
| $^{57}$Co | 122 | 477 | 873 |
| $^{60}$Co | 1332 | 1905 | 3525 |

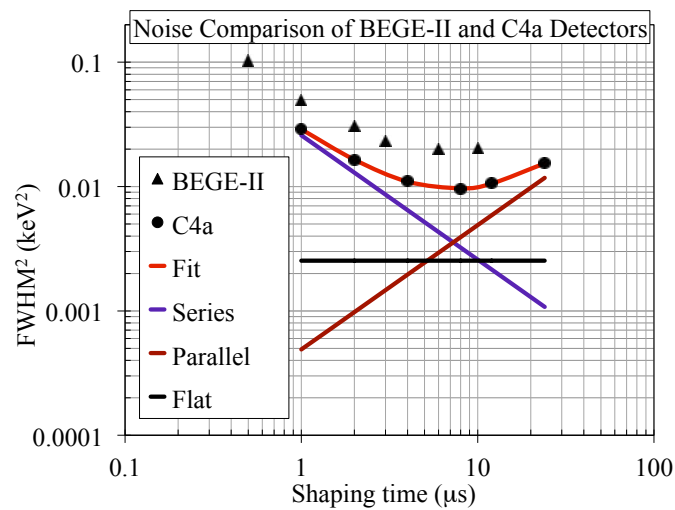

Fig. 11. Noise performance of the 90 mm p-type point contact detector ("C4a") reported in this Transaction compared to the noise performance of the CoGeNT [3] p-type point contact detector ("BEGE-II").

### C. Detector Dead-layer Characterization

The detector and cryostat described above were transported to the underground laboratory at the Kavli Institute at the University of Chicago in Chicago, IL. The dead layer thickness of the detector was measured using an approach previously employed by the GERDA and MAJORANA Collaborations [17,18]. Two dedicated spectra were taken, one with a $^{241}$Am source and one with a $^{133}$Ba source. The sources were located at distances $d_{Am}$ = 14.5 cm and $d_{Ba}$ = 39.25 cm away from the detector end cap ensuring negligible dead time is introduced due to throughput limitations of the data acquisition system, schematic shown in Fig. 12.

The detector high voltage was provided by a CANBERRA Model 3106D power supply, whereas the preamp was powered by an ORTEC 4003. The signal output was fed into a 16-bit NI 5734 ADC, which was connected to an NI PXIe-7966R FPGA module. The host PC was an NI PXIe 8133. A trapezoidal, digital pulse shaper was implemented on the FPGA using a recursive algorithm. The total shaping time was set to 16 μs with a peaking time 8 μs and a zero length flat top, that is $t_{top} = 0$ μs. A rising edge threshold trigger was used to monitor the digitally shaped signal. The trigger position was set to 80% of the 400 μs long waveforms. The sampling rate was set to 40 MS/s. The 320 μs duration pre-trigger, signal trace-data allowed monitoring of the detector noise level as well as the baseline stability.

For both runs the ratio, $R_{source}$, between the intensities, $I(\gamma$-ray energy), of the high and low energetic peaks is calculated according to:

$$R_{Am} = \frac{I(59\ keV)}{I(99\ keV) + I(103\ keV)}$$

$$R_{Ba} = \frac{I(80\ keV) + I(81\ keV)}{I(356\ keV)}$$

These ratios were also predicted for several different dead layer thicknesses by Monte Carlo radiation transport simulations of the experimental setup using the MCNPX-Polimi-2 software package [19]. The resulting ratio vs. dead layer thickness is shown in solid blue in Fig. 13. The predictions from the simulation include a systematic uncertainty in the source location of ±1 cm for the $^{241}$Am source and ±2 cm for the $^{133}$Ba source as well as a ±50 μm uncertainty in the copper end cap thickness of the detector cryostat, shown as dashed blue lines above and below the predicted central value of the ratio as a function of dead layer thickness.

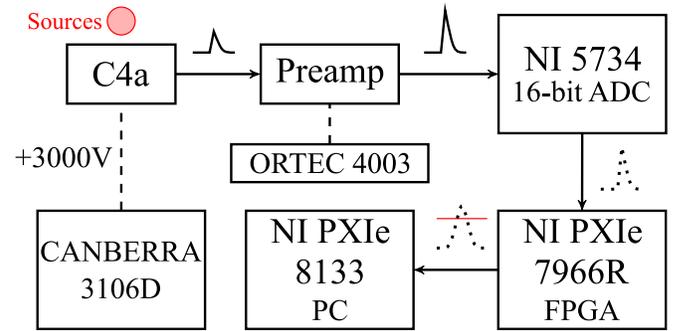

Fig. 12. Schematic of the data acquisition system used in the dead-layer measurement of the detector described in this Transaction (*i.e.*, "C4a").

Table VI shows the measured experimental ratio values from the detector's peak rates for each isotope and the extracted value of the dead layer thickness estimated from the Monte Carlo predictions presented in Fig. 13. Both measurements agree within their respective errors. The $^{241}$Am measurement favors a slightly larger dead layer, but is much more affected by the systematic uncertainties than the $^{133}$Ba data. Combining both measurements following the prescriptions of Ref. [20], the most probable dead layer thickness of $0.68^{+0.17}_{-0.12}$ mm is estimated from these measurements, in agreement with the manufacturer reported value of 0.7 mm, as reported in Table II.

TABLE VI
EXTRACTED DEAD LAYER THICKNESS

| Isotope | Experimental Ratio | Dead Layer Thickness (mm) |
|---|---|---|
| $^{241}$Am | 10.33 ± 0.64 | $0.80^{+0.24}_{-0.29}$ |
| $^{133}$Ba | 0.2534 ± 0.0030 | $0.559^{+0.041}_{-0.040}$ |

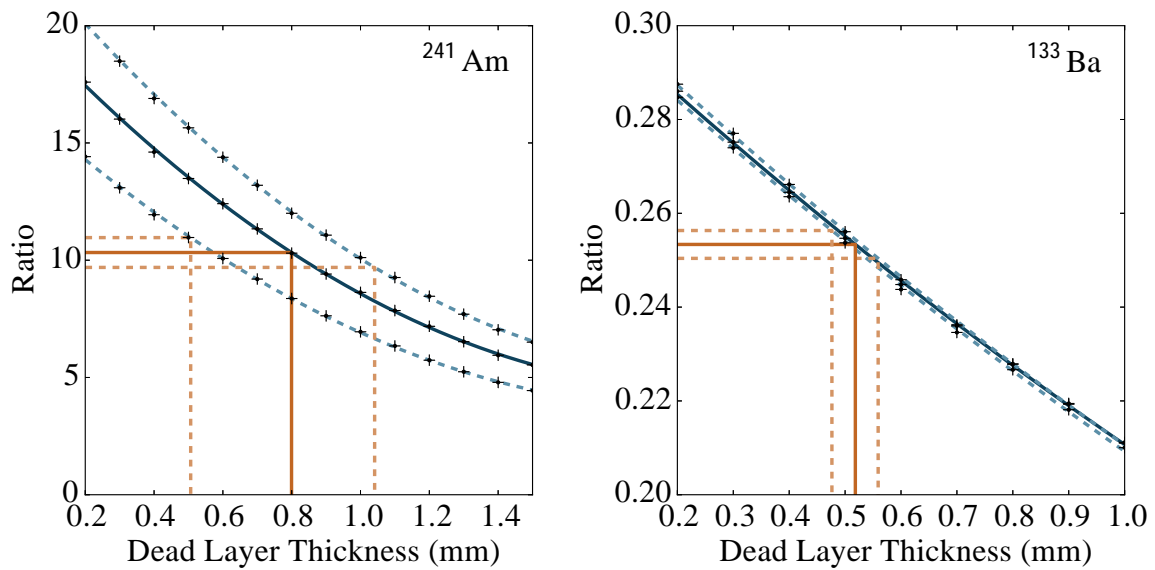

Fig. 13. A MCNPX-Polimi-2 radiation transport simulation of the detector and cryostat was performed to predict the γ-ray peak intensity ratios, $R_{Am}$ and $R_{Ba}$ (as defined in the text), as a function of the dead layer thickness of the p-type point contact detector shown in Fig. 6. The predicted central values of these ratios are shown as the solid blue, monotonically decreasing curves. The upper and lower dashed lines represent the systematic uncertainties associated with the experimental realization of the simulation geometry. Experimental data was collected to determine the γ-ray peak intensity ratios for each isotope (horizontal solid orange lines) and were used to extract (vertical solid orange lines) an estimated dead layer thickness for the p-type point contact detector. See text for details.

## IV. Conclusion

Progress has been made on the development of both a low-background cryostat design and a low-noise p-type point contact germanium ionization spectrometer. The cryostat vacuum envelope has been assembled, tested, and found to meet design requirements. Initial thermal tests have investigated the operating temperature at the end of the dipstick cold finger as well as mechanical-joint thermal resistances. A large volume p-type point contact detector was selected and characterized prior to integration into the cryostat. The detector was successfully integrated into the cryostat and operated, meeting the desired performance expectations. Reproducible manufacture of p-type point contact germanium ionization spectrometers with crystal masses of up to 2 kg and intrinsic noise figures of ~60 eV FWHM have become possible as a result of the R&D described here. This represents an improvement by a factor ~4 in mass and ~3 in noise, with respect to the previous generation of p-type point contact germanium ionization spectrometers [3].

## Acknowledgments

We thank Eric W. Hoppe and Jason Merriman for fabrication of the cryostat end cap and infrared shield using their ultra-pure copper electroforming capability. The Ultra-Sensitive Nuclear Measurement Initiative, a Laboratory Directed Research and Development program at the Pacific Northwest National Laboratory supported the development of the low background cryostat. The National Science Foundation supported the development of the low noise germanium detector via a grant to the University of Chicago (PHYS-1003940).

Craig E. Aalseth, James E. Fast, Todd W. Hossbach, John L. Orrell, Cory T. Overman, and Brent A. VanDevender are with the Pacific Northwest National Laboratory, Richland, WA 99352, USA.

Juan I. Collar and Bjorn Scholz are with the Kavli Institute for Cosmological Physics and Enrico Fermi Institute, University of Chicago, Chicago, IL 60637, USA.

Jim Colaresi and K. Mike Yocum are with CANBERRA Industries, Meriden, CT 06450, USA.